\documentclass{iau}
\usepackage{graphicx}

\title[Understanding the Last Mile] %% give here short title %%
{Understanding the Last Mile -- Physics of the Accretion Column}

\author[Peter Kretschmar et al.]   %% give here short author list %%
{%
Peter Kretschmar$^1$,
Peter A. Becker$^2$,
Dipankar Bhattacharya$^3$,  
Isabel Caballero$^4$,  
Thomas Dauser $^5$,  
Carlo Ferrigno$^6$,      
Dmitry Klochkov$^7$,        
Ingo Kreykenbohm$^5$,       
Osamu Nishimura $^8$,    
Katja Pottschmidt$^{9,10}$,      
Richard E. Rothschild$^{11}$,
Andrea Santangelo$^7$,
Gabriele Sch\"onherr$^{12}$,  
Fritz-Walter Schwarm$^5$,
R\"udiger Staubert$^7$,
Slawomir Suchy$^7$,
Brent West$^2$,
J\"orn Wilms$^5$,
Michael Wolff$^{13}$ \and          
Kenneth Wolfram$^{13}$
}

\affiliation{%
$^1$European Space Astronomy Centre (ESA/ESAC), Science Operations Department, Villanueva de la Ca\~nada (Madrid), Spain\\
$^2$School of Physics, Astronomy, and Computational Sciences, MS 5C3,\\ 
George Mason University, 4400 University Drive, Fairfax, VA, USA\\
$^3$AIM (UMR 7158 CEA/DSM – CNRS – Université Paris Diderot) Irfu/Service d’Astrophysique, 91191 Gif-sur-Yvette, France\\
$^5$Dr. Karl Remeis-Observatory and Erlangen Centre for Astroparticle Physics, \\
Sternwartstr. 7, 96049 Bamberg, Germany\\
$^6$ISDC Data Center for Astrophysics, Universit\'e de Gen\`eve, \\
Chemin d’Ecogia 16, 1290 Versoix, Switzerland\\
$^7$Institut für Astronomie und Astrophysik, Abt. Astronomie, Universit\"at T\"ubingen, \\
Sand 1, 72076 T\"ubingen, Germany\\
$^8$Department of Electronics and Computer Science, Nagano National College of Technology, 716 Tokuma, 381-8550 Nagano, Japan\\ 
$^9$CRESST and NASA Goddard Space Flight Center, Astrophysics Science Division, Code 661, Greenbelt, MD 20771, USA\\
$^{10}$Center for Space Science and Technology, University of Maryland Baltimore County, \\
1000 Hilltop Circle, Baltimore, MD 21250,
USA\\
$^{11}$Center for Astrophysics \& Space Sciences, University of California, San Diego, \\
9500 Gilman Drive, La Jolla, CA 92093, USA\\
$^{12}$Leibniz-Institut f\"ur Astrophysik Potsdam, An der Sternwarte 16, 14482 Potsdam, Germany\\
$^{13}$Space Science Division, Naval Research Laboratory, Washington, DC, USA
}

\pubyear{2012}
\volume{290}  %% insert here IAU Symposium No.
\jname{Feeding compact objects: Accretion on all scales}
\editors{C.M. Zhang, T. Belloni, M. M\'endez \& S.N. Zhang, eds.}

\begin{document}

\maketitle

\begin{abstract}
Accreting X-ray pulsars are among the best observed objects of X-ray astronomy with a 
rich data set of observational phenomena in the spectral and timing domain. While the 
general picture for these sources is well established, the detailed physics behind the 
observed phenomena are often subject of debate.

We present recent observational, theoretical and modeling results for the structure and 
dynamics of the accretion column in these systems. Our results indicate the presence of 
different accretion regimes and possible explanations for observed variations of spectral 
features with luminosity.
\end{abstract}

\firstsection % if your document starts with a section,
\section{Introduction}

Accreting X-ray pulsars exist in binary systems, where a companion star
transfers matter to the neutron star via Roche lobe overflow, a strong stellar wind 
or a Be outflow disk, \cite[e.g., Frank, King \& Raine (2002)]{FKR:2002}. 
The sizes of these systems vary widely, generally in the range $10^{6}$ to $10^{8}$\,km.

Independently of the way matter is transferred, close to the compact object the flow of matter 
is dominated by the strong magnetic field $B\sim10^{12}$\,G. The accreting plasma couples to
the magnetic field lines at the Alv\'en radius, which is a few 1000\,km for typical neutron star 
parameters (see, e.g., \cite{CaballeroWilms:2012} and references therein). The infalling matter follows the magnetic field lines, forming accretion columns 
on the neutron star magnetic poles with a surface area of $\sim$1\,km$^2$ and a height that
can range from meters to many km.

The observed X-ray spectra are usually relatively simple and broadly similar across
sources. They are normally well described by a power law with
exponential cutoff, affected by absorption and including iron fluorescence lines which are 
produced within the binary system, as well close as far away from the X-ray source. 
In some accreting X-ray pulsars, broad absorption-like
features are observed which are caused by resonant scattering of photons on electrons whose
energies are quantized into so-called Landau levels, spaced by
$$
   E_{\rm cyc} = \hbar { e B \over m_{\rm e} c} 
   \approx 11.6\,{\rm keV} {B \over 10^{12} {\rm G}}.
$$
For an overview of known cyclotron line sources see \cite{CaballeroWilms:2012}. 

For bright accreting X-ray pulsars \cite{BeckerWolff:2007} have developed a model of the continuum spectra, where the total observed spectrum is a result of the bulk and thermal Comptonization of bremstrahlung, black body and cyclotron seed photons. This model has been adapted to 
\texttt{xspec} and used for analysis of 4U\,0115+63 by \cite{Ferrigno:2009}.

Cyclotron line features are still usually fitted phenomenologically with independent 
Gaussian or Lorentzian line shapes. A more physical approach to model
the complex harmonic cyclotron line shapes simultaneously has been pursued by
\cite{Schoenherr:2007}: Extensive Monte-Carlo simulations of the cyclotron 
resonant scattering on a large parameter grid 
(following \cite[Araya \& Harding 1999, 2000]{Araya:1999}) have been used to create
a Green's function based \texttt{xspec} convolution model, applicable to any continuum
and successfully applied to spectra of different sources, e.g., by 
\cite{Suchy:2008a}.

%These continuum and line models have been applied successfully to average spectra
%of different sources, but for
%a self-consistent interpretation of pulse-phase-resolved spectra, a more general
%approach is required (see Sect.~\ref{new}).

In contrast to the spectra, the pulse profiles -- light curves folded by the pulse 
period -- can be quite unique for a given source, with the typical shape serving
almost as a `fingerprint'. The profiles are usually quite dependent on the energy
band, with a common tendency to have simpler profiles at higher energies. 
 In some sources these energy-dependent profiles are very stable across
significant brightness variations, in other cases a clear evolution is visible and
occasionally rather chaotic changes have been observed 
\cite[(e.g., Camero-Arranz \etal\ 2007)]{Camero-Arranz:2007}. 
What drives these different modes of behaviour is still not well understood.

\section{Recent Results}

\subsection{Pulse profile modeling}
\cite{Caballero:2011} and \cite{Sasaki:2012} have applied methods developed
by \cite{Kraus:95} in order to decompose the pulse profiles of the sources
\mbox{A\,0535+26}, \mbox{4U\,0115+63} and \mbox{V\,0332+53}. In this approach
axisymmetrical emission and the same beam pattern is assumed for both poles,
but the poles can be offset from antipodal. The observed profiles can be explained
within these models including emission from the accretion column a halo around 
the footpoint of the column as well as scattering of photons in the upper accretion 
stream. The change of emission geometry during a giant outburst of \mbox{V\,0332+53}
could be demonstrated by \cite{Sasaki:2012}.

\subsection{Cyclotron line energy variations}
For some of the sources with cyclotron lines the derived line energies have been found
to vary as a function of luminosity, see, e.g., \cite{Mowlavi:2006}, \cite{Tsygankov:2007},
\cite{Staubert:2007}, or \cite{Klochkov:2011a}. The data point towards two types of variability: 
for certain sources, (e.g., V\,0332+53) the centroid energy correlates \emph{negatively} with the luminosity, for others (e.g., Her~X-1) a \emph{positive} correlation is observed, while for some
sources (e.g., A\,0535+26) the line appears to be at constant energies (within uncertainties) despite significant luminosity changes. A possible explanation has been developed by
\cite{Becker:2012} who describe different accretion regimes based on the definition of 
a critical luminosity $L_{\rm crit} = 1.5\times 10^{37} (B/10^{12}\,G) $ above which radiation pressure decelerates the matter to rest. The brightest sources ("supercritical", i.e., 
$L_{\rm X}>L_{\rm crit}$), will have an accretion column where the characteristic emission height increases with luminosity and the observed $B$ field decreases accordingly. In subcritical 
sources, somewhat below $L_{\rm crit}$, Coulomb braking will be dominant and increased luminosity leads
to a decreasing emission height and increasing observed $B$ field. At even lower luminosities no
significant evolution is expected. This model agrees well with the observed source behaviour and
parameters, for details see \cite{Becker:2012}.

\subsection{Accretion mound stability and variations with pulse phase}

A common feature of cyclotron line sources is that the observed line energy varies with pulse phase. While a variation with angle is expected if one includes relativistic corrections, the observed variations are often larger than those predicted straightforwardly.

\cite{MukherjeeBhattacharya:2012} and colleagues have studied the stability of accretion
mounds and the distribution of the magnetic field in the mound. For mounds of a total accreted mass of $\sim 10^{-12}M_{\odot}$ there is an appreciable distortion of the magnetic field
with a stronger relative field towards the rim of the mound. This distortion would have a strong signature in the observed spectra if mound surface were visible directly.
In the 2D MHD simulations ballooning instabilities develop which limit the maximum height of the
mound and the amount of matter that can be contained. Beyond $\sim 10^{-13}M_{\odot}$ 
the mounds become unstable.

\section{New developments}\label{new}
In order to progress significantly beyond the efforts described above, a challenging overall problem needs to be tackled:
For each pole the X-ray continuum production needs to be modeled as function of energy
and position in the column. This emission is further significantly modified by cyclotron
resonant scattering strongly dependent on energy and angle of photon propagation. To
describe the observed spectra or pulse profiles the emission from both poles must be
combined using realistic emission geometries and including relativistic light bending
around the neutron star.

As one step in this process, members of our collaboration are further developing the 
Green's function models of cyclotron line scattering processes (\cite[Sch\"onherr \etal\ \textit{in prep.}]{Schoenherr:COSPAR2012}, \cite[Schwarm \etal\ \textit{in prep.}]{Schwarm:IWS2012}). Compared to previous
approaches (e.g., \cite[Araya \& Harding 1999, 2000]{Araya:1999}) much more general
geometries are being implemented, allowing the adaption to light-bending
models for the effectively visible emission. This work is cross-checked
with other approaches, see, e.g., \cite[Nishimura (2008, 2011)]{Nishimura:2008}.

In parallel an effort is ongoing to combine the modeling efforts of \cite{BeckerWolff:2007}
and \cite{Becker:2012} with the advanced cyclotron line scattering description. 
In this `hybrid' approach, column properties and seed photons are derived from the continuum 
models, which get modified in an outer 'sheath' by the advanced cyclotron scattering models.
An additional component in this effort is the inclusion of ray tracing within a general 
relativistic geometry in order to describe the corresponding modifications to the observed
spectra.

The structure of accretion mounds is further studied in 3D MHD simulations by 
\cite[Mukherje, Bhattacharya \& Mignone (\textit{in prep.})]{MukherjeeBhattacharyaMignone:201X}.
First results indicate an easy excitation of the fluting mode instability and consequent transport
of matter across the field. This would reduce the maximum mass supported within the mounds
and field distortions would be on small scales.

\section{Summary}
In recent years significant progress has been made on various fronts in the understanding of
the physics of accretion columns in X-ray pulsars. These developments have
been possible due to improved methods and advances in computing power. The results
are encouraging, but also demonstrate the complexity of the issues which have to be taken
into account.

Within our collaboration work is now ongoing to improve the existing models and to
combine different approaches in order to provide detailed predictions for direct
comparison with the rich observational dataset.

\begin{acknowledgements}
This work has been performed within the MAGNET collaboration. We would like to thank the ISSI, Bern, for hosting stimulating and productive international team meetings focussed on the work presented here.
\end{acknowledgements}

\end{document}